\documentstyle[12pt]{article}
\catcode`\@=11

\@addtoreset{equation}{section}
\def\theequation{\arabic{section}.\arabic{equation}}

\def\@normalsize{\@setsize\normalsize{15pt}\xiipt\@xiipt
\abovedisplayskip 14pt plus3pt minus3pt%
\belowdisplayskip \abovedisplayskip
\abovedisplayshortskip  \z@ plus3pt%
\belowdisplayshortskip  7pt plus3.5pt minus0pt}

\def\small{\@setsize\small{13.6pt}\xipt\@xipt
\abovedisplayskip 13pt plus3pt minus3pt%
\belowdisplayskip \abovedisplayskip
\abovedisplayshortskip  \z@ plus3pt%
\belowdisplayshortskip  7pt plus3.5pt minus0pt
\def\@listi{\parsep 4.5pt plus 2pt minus 1pt
            \itemsep \parsep
            \topsep 9pt plus 3pt minus 3pt}}

\def\underline#1{\relax\ifmmode\@@underline#1\else
        $\@@underline{\hbox{#1}}$\relax\fi}
\@twosidetrue
\relax

\catcode`@=12

\catcode`\@=11

\def\section{\@startsection{section}{1}{\z@}{3.5ex plus 1ex minus
   .2ex}{2.3ex plus .2ex}{\large\bf}}
\def\thesection{\arabic{section}.}

\def\ps@headings{\def\@oddfoot{}\def\@evenfoot{}
\def\@oddhead{\hbox{}\hfill
        \makebox[.5\textwidth]{\raggedright\ignorespaces --\thepage{}--
        \hfill }}
\def\@evenhead{\@oddhead}
\def\subsectionmark##1{\markboth{##1}{}}
}

\ps@headings

\catcode`\@=12

\relax

\def\figcap{\section*{Figure Captions\markboth
        {FIGURECAPTIONS}{FIGURECAPTIONS}}\list
        {Fig. \arabic{enumi}:\hfill}{\settowidth\labelwidth{Fig. 999:}
        \leftmargin\labelwidth
        \advance\leftmargin\labelsep\usecounter{enumi}}}
 \relax
\def\tablecap{\section*{Table Captions\markboth
        {TABLECAPTIONS}{TABLECAPTIONS}}\list
        {Table \arabic{enumi}:\hfill}{\settowidth\labelwidth{Table 999:}
        \leftmargin\labelwidth
        \advance\leftmargin\labelsep\usecounter{enumi}}}
 \relax
\def\reflist{\section*{References\markboth
        {REFLIST}{REFLIST}}\list
        {[\arabic{enumi}]\hfill}{\settowidth\labelwidth{[999]}
        \leftmargin\labelwidth
        \advance\leftmargin\labelsep\usecounter{enumi}}}
 \relax

\catcode`\@=11

\def\ps@headings{\def\@oddfoot{}\def\@evenfoot{}
\def\@oddhead{\hbox{}\hfill
        \makebox[.5\textwidth]{\raggedright\ignorespaces --\thepage{}--
        \hfill }}
\def\@evenhead{\@oddhead}
\def\subsectionmark##1{\markboth{##1}{}}
}

\ps@headings

\relax

\def\firstpage#1#2#3#4#5#6{
\begin{document}
\begin{titlepage}
\nopagebreak
\title{\begin{flushright}
        \vspace*{-1.5in}
        {\normalsize NUB--#1\\[-3mm]
        #2\\[-9mm]CPTH--A185.0892}\\[6mm]
\end{flushright}
\vfill
{\large \bf #3}}
\author{\large #4 \\[1cm] #5}
\maketitle
\vfill
\nopagebreak
\begin{abstract}
{\noindent #6}
\end{abstract}
\vfill
\begin{flushleft}
\rule{16.1cm}{0.2mm}\\[-3mm]
$^{\star}${\small Work supported in part by the \vspace{-4mm}
Northeastern University Research and Scholarship Development
Fund, in part by the National Science Foundation \vspace{-4mm}
under grant PHY--91--07809,  and in part by the EEC
contracts SC1--0394--C, SC1--915053 and SC1--CT92--0792.}\\
September 1992
\end{flushleft}
\thispagestyle{empty}
\end{titlepage}}
\newcommand{\dal}{\raisebox{0.085cm}
{\fbox{\rule{0cm}{0.07cm}\,}}}
\newcommand{\gef}{G_{\rm eff}}
\newcommand{\lef}{\Lambda_{\rm eff}}
\newcommand{\dt}{\partial_{\langle T\rangle}}
\newcommand{\dtbar}{\partial_{\langle\overline{T}\rangle}}
\newcommand{\al}{\alpha^{\prime}}
\newcommand{\mst}{M_{\scriptscriptstyle \!S}}
\newcommand{\mpl}{M_{\scriptscriptstyle \!P}}
\newcommand{\alg}{\alpha_{\scriptscriptstyle G}}
\newcommand{\bg}{\beta_{\scriptscriptstyle G}}
\newcommand{\bl}{\beta_{\scriptscriptstyle \Lambda}}
\newcommand{\gn}{\frac{1}{16\pi G}}
\newcommand{\gr}{G_{\!\rm eff}}
\newcommand{\eff}{\Gamma_{\!\rm eff}}
\newcommand{\lr}{\Lambda_{\rm eff}}
\newcommand{\gnn}{\frac{1}{32\pi G}}
\newcommand{\dv}{\int{\rm d}^4x\sqrt{g}}
\newcommand{\lac}{\lambda_{\scriptscriptstyle G}}
\newcommand{\act}{\widetilde{\Gamma}}
\newcommand{\lv}{\left\langle}
\newcommand{\rv}{\right\rangle}
\newcommand{\ph}{\varphi}
\newcommand{\sbar}{\,\overline{\! S}}
\newcommand{\xbar}{\,\overline{\! X}}
\newcommand{\barz}{\,\overline{\! Z}}
\newcommand{\zbar}{\bar{z}}
\newcommand{\dbar}{\,\overline{\!\partial}}
\newcommand{\tbar}{\overline{T}}
\newcommand{\psibar}{\overline{\Psi}}
\newcommand{\ybar}{\overline{Y}}
\newcommand{\z}{\zeta}
\newcommand{\zb}{\bar{\zeta}}
\newcommand{\phb}{\overline{\varphi}}
\newcommand{\cm}{Commun.\ Math.\ Phys.~}
\newcommand{\pr}{Phys.\ Rev.\ D~}
\newcommand{\pl}{Phys.\ Lett.\ B~}
\newcommand{\np}{Nucl.\ Phys.\ B~}
\newcommand{\e}{{\rm e}}
\newcommand{\gsi}{\,\raisebox{-0.13cm}{$\stackrel{\textstyle
>}{\textstyle\sim}$}\,}
\newcommand{\lsi}{\,\raisebox{-0.13cm}{$\stackrel{\textstyle
<}{\textstyle\sim}$}\,}
\date{}
\firstpage{3057}{IC/92/402}
{\large\sc Superstring Threshold Corrections to Yukawa
Couplings$^\star$}
{I. Antoniadis$^{\,a}$, E. Gava$^{b,c}$,
K.S. Narain$^{ c}$ $\,$and$\,$
T.R. Taylor$^{\,d}$}
{\normalsize\sl
$^a$Centre de Physique Th\'eorique, Ecole Polytechnique,
F-91128 Palaiseau, France\\
\normalsize\sl
$^b$Instituto Nazionale di Fisica Nucleare, sez.\ di Trieste,
Italy\\
\normalsize\sl $^c$International Centre for Theoretical Physics,
I-34100 Trieste, Italy\\
\normalsize\sl $^d$Department of Physics, Northeastern
University, Boston, MA 02115, U.S.A.}
{A general method of computing string corrections to the K\"ahler
metric and Yukawa couplings is developed at the one-loop level for
a general compactification of the heterotic superstring theory. It
also provides a direct determination of the so-called Green-Schwarz
term. The matter metric has an infrared divergent part which
reproduces the field-theoretical anomalous dimensions, and a
moduli-dependent part which gives rise to threshold corrections in
the physical Yukawa couplings. Explicit expressions are derived for
symmetric orbifold compactifications.}

\setcounter{section}{0}
\section{Introduction}
A fundamental task for the heterotic superstring theory is the
determination of the effective action describing the physics of
massless string excitations at low energies. This is necessary for
the phenomenological applications of string theory, in particular
for the unification of gauge interactions. The most general
$N=1$ supergravity action, describing local interactions
involving up to two space-time derivatives, is characterized by
three functions of chiral superfields: the real K\"ahler potential
$K$ which determines the kinetic terms, the analytic
superpotential $W$ related to the Yukawa couplings, and the
analytic function $f$ associated with the gauge couplings \cite{cet}. From
the
phenomenological point of view, the most important issue is the dependence
of
these functions on the moduli-fields. The reason is that the vacuum
expectation
values (VEVs) of the moduli, which are completely arbitrary in perturbation
theory, may drastically affect the values of the observable couplings.
This moduli-dependence is restricted by the space-time duality
symmetry of the four-dimensional heterotic string theory.

The gauge couplings do not depend on the moduli at the tree-level --
they are determined by the dilaton VEV. The moduli-dependence
of the radiative corrections to gauge couplings satisfies a
non-renormalization theorem \cite{ant}, namely, it is given entirely by
the one loop contributions \cite{dkl2,ant}. The gauge $f$-functions can
be determined by using duality symmetry, from the violation of the
integrability condition of the corresponding $\Theta$-angles.
The gauge group-dependent part of the latter is due
to the presence of anomalous diagrams involving massless particles,
and can be calculated at the level of the effective field theory
\cite{dfkz,anom}.
The $f$-functions as well as the related string threshold
corrections have been explicitly determined for the symmetric
orbifold compactifications, and the general procedure
is very clear.

The physical Yukawa couplings depend on the superpotential $W$, as
well as on the K\"ahler potential $K$ through wave functions. At the
tree level, both $W$ and $K$ are known in many cases (including
symmetric orbifolds), up to the terms trilinear and quadratic in
matter fields, respectively. In (2,2) models, the $N=2$ world-sheet
supersymmetry of the bosonic sector implies strong tree-level
relations  between these two functions \cite{dkl1}. In particular, the
K\"ahler manifold of moduli-fields exhibits the so-called special
geometry \cite{sg,dkl1}, which relates the moduli K\"ahler potential
with the matter superpotential.

In this work we study the moduli-dependence of the radiative
corrections to physical Yukawa couplings. As a consequence of
non-renormalization theorems, the superpotential does not
receive any loop corrections \cite{m}. Hence it is sufficient to
consider the higher genus string contributions to the K\"ahler
metric. In section 2, we describe the general method
of the computation of matter and moduli metrics and
we obtain the general one-loop expression. The method is based  on
the evaluation of the three-point amplitude involving the two-index
antisymmetric tensor and two scalar fields. In section 3, we show
that the above expression coincides with the universal part of the
violation of the integrability condition for $\Theta$-angles. This
result is not surprising: in fact our method provides a direct
string evaluation of the so-called Green-Schwarz term
\cite{gs,dfkz}. We also point out a connection to the ${\rm
Tr}F(-1)^F$ ``index" of the underlying $N=2$ superconformal theory
\cite{cv}. In section 4, we analyze the infrared divergencies present
in the matter metric and we identify the corresponding contribution to
the field-theoretical one-loop beta functions of the Yukawa couplings.
We also define the threshold corrections. In sections 5 and 6, we
consider the example of symmetric orbifolds. In section 5, we derive
explicit expressions for the one-loop metric of the untwisted moduli,
and we point out that special geometry is violated beyond the tree
approximation. In section 6, we derive the one-loop metric for the
untwisted matter fields and we determine the threshold corrections to
their Yukawa couplings.

\section{One-loop K\"ahler metric and physical Yukawa couplings}

In the standard formulation of $N=1$ supergravity \cite{cet}, all
massless scalars and pseudoscalars form complex fields $z$ which
together with their fermionic partners $\chi$ belong to
chiral supermultiplets $Z$. They parametrize a K\"ahler manifold
of a non-linear sigma-model with the metric
$K_{i\bar{\jmath}}=\partial_{z^i}\partial_{\bar{z}^{\bar{\jmath}}}
K(z,\bar{z})$. At the tree-level, the K\"ahler metric can be extracted
from string scattering amplitudes involving four or more complex scalar
fields \cite{dkl1}. Such a computation becomes very complicated beyond
the tree approximation. On the other hand, direct computation of
three-point Yukawa couplings requires fermion vertices which are quite
difficult to handle. Another complication is due to the
existence of the dilaton which belongs to a very distinct
supersymmetry multiplet, together with the two-index
antisymmetric tensor which is equivalent to a pseudoscalar axion.
Since the dilaton VEV plays the role of the string loop
expansion parameter, string loop corrections give rise to
kinetic terms that mix the moduli with the dilaton and axion.
This complicates the ana\-ly\-sis of the moduli-dependence of the
K\"ahler potential. As we explain below, the simplest way to avoid all
these problems is to represent the dilaton supermultiplet by a
real linear superfield $L$, satisfying the constraint ${\cal D}^2
L={\bar{\cal D}}^2 L=0$, in the rigid supersymmetry notation. The
linear representation makes direct use of the antisymmetric tensor,
which is natural in view of the form of the corresponding string vertex
operator.

In the linear formulation, the kinetic terms
originate, up to one-loop order, from the $D$-density of the function
\cite{fv}:
\begin{equation}
{\cal L}~=~-
\left(\frac{L}{2}\right)^{-1/2}\!(\Sigma{\overline \Sigma})^{3/2}\,
e^{\frac{1}{2}G^{(0)}(Z,\barz)} ~+~
\frac{L}{2} G^{(1)}(Z,\barz),
\label{dterm}
\end{equation}
where $\Sigma$ is the so-called chiral compensator field
\cite{cet,ku}, and $G^{(0)}$, $G^{(1)}$ are real functions of chiral
superfields.\footnote{Here, we do not discuss the $z$-independent
corrections to the dilaton kinetic term.} As usual, the chiral
compensator is fixed by normalizing the gravitational kinetic energy
term to $-\frac{1}{2}R$. This reduces to the following condition for
the scalar components:
\begin{equation}
\sigma^3 =
\sqrt{l/2}\,e^{\frac{1}{2}G^{(0)}(z,\bar{z})},
\label{comp}
\end{equation}
where $l$ is the exponential of the dilaton, so that its VEV determines
the four-dimensional string coupling constant, $\lv l\,\rv =g^2$. Note
that the chiral compensator $\sigma$ does not depend on the function
$G^{(1)}$. With this normalization, the bosonic part of the kinetic
energy terms is given by:
\begin{equation}
{\cal L}_{\rm b} ~=~
- \frac{1}{4l^2}\partial_{\mu}l\partial^{\mu}l + \frac{1}{4l^2}
h_{\mu}h^{\mu} - G_{i\bar{\jmath}}\partial_{\mu}{z^i}\partial^{\mu}
{\bar{z}^{\bar{\jmath}}} -
\frac{i}{2}(G_{lj}\partial_{\mu}{z^j} -
G_{l\bar{\jmath}}\partial_{\mu}{\bar{z}^{\bar{\jmath}}})h^{\mu},
\label{LB}
\end{equation}
where
\begin{equation}
h^{\mu}=\frac{1}{2}\epsilon^{\mu\nu\lambda\rho}
\partial_{\nu}b_{\lambda\rho}
\label{Hmu}
\end{equation}
is the dual field strength of the
antisymmetric tensor $b_{\lambda\rho}$. The function $G$, which
determines the K\"ahler metric, is:
\begin{equation}
G(l,z,\zbar) = G^{(0)}(z,\zbar) + l G^{(1)}(z,\zbar).
\label{G}
\end{equation}
In eq.(\ref{LB}) the subscripts of $G$ denote differentiations with
respect to the corresponding fields.

To relate the linear to the standard formulation,
in which the dilaton multiplet is represented by a chiral superfield
$S$, one introduces the latter as a Lagrange multiplier and
rewrites the Lagrangian as a $D$-density of
${\cal L} - L(S + \sbar)/4$ \cite{fv,dfkz}. The equation of motion
for $S$ imposes the linearity constraint for $L$. On the other hand, if
one uses first the equation of motion for $L$, one obtains
$L(S,\sbar,Z,\barz,\Sigma,\overline{\Sigma})$ as a solution of the
equation $\partial_L{\cal L} = (S + \sbar)/4$. The Lagrangian is
then given by the $D$-density of ${\cal L} - L \partial_L{\cal L}$
which must be identified with the standard Lagrangian $-\frac{3}{2}
\Sigma\overline{\Sigma}\,e^{-\frac{1}{3}K(S,\sbar,Z,\barz)}$. In
this way one obtains $K=-\ln [S+\sbar -2G^{(1)}]+G^{(0)}$. It is now
obvious that the presence of moduli-dependent correction $G^{(1)}$
induces kinetic terms which mix $S$ with the  moduli. The linear
formulation provides therefore a very convenient field basis,
in which the dilaton does not mix with any other field even in the
presence of  non-trivial loop corrections, see eq.(\ref{LB}).

The Yukawa interactions originate, in the linear formulation, from the
field-dependent fermion mass terms contained in the $F$-density of
the function $\Sigma^3 W(Z)$. $W$ is the analytic superpotential, which
cannot depend on $L$ and therefore it is completely determined at the
tree-level, as expected from supersymmetric non-renormalization
theorems. With the chiral compensator normalized as in (\ref{comp}),
the Yukawa interactions between massless matter fields are:
\begin{equation}
{\cal L}_Y~=~-\frac{1}{2}\sqrt{l/2}\,e^{G^{(0)}/2}\,W_{ijk}\,
\chi^i \chi^j z^k ~+~ h.c.
\label{ly}
\end{equation}
where the subscripts of $W$ denote differentiations with respect to
the corresponding fields. Here, we assumed unbroken supersymmetry with
$W=\partial_z W=0$ in the vacuum. Note that the above expression
depends on the tree-level quantities only. The physical Yukawa
couplings, defined by the fermion-fermion-scalar S-matrix elements may
receive however loop corrections. They arise from the corrections to
the K\"ahler metric [see (\ref{G})] which change the wave function
normalization factors.

For our purposes, the main advantage of using the linear formulation
is that it provides a simple way of computing the loop corrections
to the K\"ahler metric, by considering the three-point amplitude
involving two complex scalars and the antisymmetric
tensor. Inspection of the last term in eq.(\ref{LB}) shows that
$G^{(1)}_{z\zbar }$ can be determined from the $C\!P$-odd part of the
correlation function:
\begin{equation}
\langle z(p_1)\zbar (p_2)b^{\mu\nu}(p_3)\rangle_{\rm odd} ~=~
i\epsilon^{\mu\nu\lambda\rho}p_{1\lambda}p_{2\rho}\,
G^{(1)}_{z\zbar},
\label{amp}
\end{equation}
where $p_1$, $p_2$ and $p_3$ are the corresponding external
momenta. Although this amplitude vanishes for on-shell Minkowski
momenta, it can be computed for complex Euclidean momenta, as it
was done in similar computations for moduli and gauge bosons
\cite{dkl2}.

In the superstring computation, the amplitude (\ref{amp}) receives
contributions only from the odd spin structures. The one-loop
contribution to the K\"ahler metric is given by:
\begin{equation}
i\epsilon^{\mu\nu\lambda\rho}p_{1\lambda}p_{2\rho}\, G^{(1)}_{z\zbar}
{}~=~\int_{\Gamma}\frac{d^2\tau}{{\tau_2}}
\int\prod_{i=1}^{3} [d^2\z_i]
\left\langle V_z(p_1,\z_1)\, V_{\zbar} (p_2,\z_2)\,
V_b^{(-1)\mu\nu}(p_3,\z_3)\, {\cal T}_F(0)\right\rangle_{\rm odd}
\label{zzb}
\end{equation}
where $\tau = \tau_1 + i\tau_2$ is the Teichm\"uller parameter of the
world-sheet torus and  ${\Gamma}$ its fundamental domain.
The vertex operators are
\begin{equation}
V_z(p,\z) ~=~ :\!(\Phi_z + ip\!\cdot\!\psi\,\Psi_z) e^{ip\cdot
\!X}\!:~,\label{vz}\vspace{-3mm}
\end{equation}\begin{equation}
V_{\zbar}(p,\z) ~=~ :\!(\overline{\Phi}_{\zbar} +
ip\!\cdot\!\psi\,\psibar_{\zbar}) e^{ip\cdot \!X}\!:~,\label{vzb}
\vspace{1mm}
\end{equation}\begin{equation}
V_b^{(-1)\mu\nu}(p,\z) ~=~ :\! i\dbar X^{[\mu}\, \psi^{\nu ]}
e^{ip\cdot \!X}\!:~,\label{vb}
\end{equation}
where the square brackets in the last equation denote
antisymmetrization in $\mu,\nu$. $X^{\mu}$ represent the
bosonic space-time coordinates and $\psi^{\mu}$ are their
left-moving fermionic superpartners. The complex space-time scalars
$z$ $(\zbar)$  correspond, in the underlying $N=2$ internal
superconformal theory, to chiral (anti-chiral) $N=2$
supermultiplets. Their lower  components are primary fields $\Psi_z$
$(\psibar_{\zbar})$ having dimensions $(\frac{1}{2},1)$ while their
upper components $\Phi_z$ $(\overline{\Phi}_{\zbar})$ have
dimensions $(1,1)$ and are given by $\Phi_z = \frac{1}{2\pi i}
\oint 2 {\cal T}_F{\Psi_z}$. The primary fields $\Psi_z$, $\psibar_{\zbar}$
define the K\"ahler metric by their two-point function
$\Psi_z(\z)\psibar_{\zbar}(0) \sim G^{(0)}_{z\zbar}/{\z{\bar{\z}}^2}$.
Finally, the supercurrent
${\cal T}_F$ insertion and the $(-1)$ ghost picture for the
antisymmetric tensor vertex are due to the odd spin-structure of
the amplitude \cite{fms}. The supercurrent is given by
\begin{equation}
{\cal T}_F ~=~:\!\psi^{\alpha}\partial X_{\alpha}\!: +
\dots \, ,
\label{tf}\end{equation}
where we omitted the part corresponding to the internal
superconformal theory, as well as the ghost part.

The four space-time zero-modes required in the odd spin structure
give rise to the kinematic factor
$\epsilon^{\nu\alpha\lambda\rho}p_{1\lambda}p_{2\rho}$; hence
we can set $p_1=p_2=p_3=0$ everywhere else. The contraction of
$\dbar X^{\mu}$ from the antisymmetric tensor vertex (\ref{vb})
with $\partial X_{\alpha}$ from the supercurrent (\ref{tf}) gives
$\langle \dbar X^{\mu}(\zb_3) \partial X_{\alpha}(0)\rangle =
-\delta^{\mu}_{\alpha}{\pi}/{4\tau_2}$. These terms combine to yield the
kinematical factor
$\epsilon^{\mu\nu\lambda\rho}p_{1\lambda}p_{2\rho}$ in the
amplitude (\ref{amp}). After performing the $\z_2$ and $\z_3$
integrations and taking into account all the normalizations, we find:
\begin{equation}
G^{(1)}_{z\zbar}
=\frac{1}{8{(2\pi)}^5}\int_{\Gamma}\frac{d^2\tau}{{\tau_2}^2} \int
d^2{\z} \,\bar{\eta}(\bar{\tau})^{-2}\, \langle\Psi_z(\z)\psibar_{\zbar}(0)
\rangle_{\rm odd}\, ,
\label{gzz}
\end{equation}
where $\eta$ is the Dedekind eta function, and the correlation
function is computed in the internal superconformal theory. The
general expression (\ref{gzz}) allows the determination of the
one-loop K\"ahler metric for any four-dimensional heterotic
superstring model.
\setcounter{equation}{0}

\section{Green--Schwarz term}

In this section we will relate the one-loop correction to the
K\"ahler metric with the universal part of the violation of the
integrability condition for $\Theta$-angles. In the linear
formulation, there are two sources of gauge kinetic terms. (i)
The $F$-term of $\frac{1}{4}f(Z)W^\alpha W_\alpha$, where $f$ is an
analytic function of chiral superfields and $W^\alpha$ is the  gauge
field-strength superfield \cite{cet}. (ii) In the $D$-density of the
function ${\cal L}$ in (\ref{dterm}), the linear superfield $L$ can be
replaced by $L-2k\Omega$, where $\Omega$ is the Chern-Simons (real)
superfield $({\bar{\cal D}}^2\Omega=W^\alpha W_\alpha )$ \cite{fv}, and
$k$ is an arbitrary constant. This generates, in the component
notation, a gauge kinetic term of the form $-\frac{k}{4}[\frac{1}{l} +
G^{(1)}] F_{\mu\nu}F^{\mu\nu}$. Moreover, in the bosonic Lagrangian
(\ref{LB}), $h^{\mu}$ is replaced by
$h^{\mu}-\frac{k}{2}\omega^{\mu}$, where $\omega^{\mu}$ is the gauge
topological current
$(\partial_{\mu}\omega^{\mu}=F_{\mu\nu}\widetilde{F}^{\mu\nu})$.
Although $\omega^{\mu}$ is not gauge invariant, the invariance of the
Lagrangian is ensured by the appropriate tranformation property of the
antisymmetric tensor, which leaves the combination
$h^{\mu}-\frac{k}{2}\omega^{\mu}$ inert.

In heterotic superstring theory, the tree-level coupling of the
dilaton dictates the constant $k$ to be equal to the Ka\v{c}-Moody
level \cite{gins}. The analytic function $f$
vanishes at the tree-level and receives corrections only at the
one-loop \cite{ant}. The bosonic Lagrangian terms bilinear in gauge fields
are
contained in:
\begin{equation}
{\cal L}_{\rm g}=-\frac{1}{4}\Delta \, F_{\mu\nu}F^{\mu\nu}
+\frac{ik}{4}(G^{(1)}_{j}\partial_{\mu}{z^j}-
G^{(1)}_{\bar{\jmath}}\partial_{\mu}{\bar{z}^{\bar{\jmath}}})
\omega^{\mu} -\frac{k}{4l^2}h_{\mu}\omega^{\mu}
+\frac{1}{4}{\rm Im}f F_{\mu\nu}\widetilde{F}^{\mu\nu},
\label{ff}
\end{equation}
where
\begin{equation}
\Delta = k [\frac{1}{l} + G^{(1)}] + {\rm Re}f
\label{del}
\end{equation}
corresponds to the inverse square of the gauge coupling.
The second term in (\ref{ff}) is called Green-Schwarz term because it
can be interpreted as the compactification of the ten-dimensional
Chern-Simons term involved in the Green-Schwarz anomaly cancellation
mechanism \cite{gs}.

The first term in (\ref{del}) shows that if
$G^{(1)}_{z\zbar}\neq 0$, {\em i.e}.\ if the K\"ahler metric
$G_{z\zbar}$ receives loop corrections, the gauge couplings become
non-harmonic functions of complex fields,
$\Delta_{z\zbar}=kG^{(1)}_{z\zbar}$. In this case, the pseudoscalar
(axionic) couplings to gauge bosons arising from the Green-Schwarz
term violate the integrability condition for the corresponding
$\Theta$-angles:
\begin{equation}
\frac{i}{2}\,(\partial_{\zbar}\Theta_z - \partial_z\Theta_{\zbar})
=\Delta_{z\zbar}\neq 0 ,
\label{int}
\end{equation}
where the axionic coupling $\Theta_z$ of the pseudoscalar ${\rm Im}z$
to two gauge bosons is defined by the $C\! P$-odd part of the
amplitude:\footnote{For our purposes, we consider only the case in
which $z$ is neutral with respect to the gauge \linebreak bosons
$A^{\mu ,\nu}$.}
\begin{equation}
\langle A^{\mu}(p_1)A^{\nu}(p_2)z(p_3)\rangle_{\rm odd} ~=~
i\epsilon^{\mu\nu\lambda\rho}p_{1\lambda}p_{2\rho}\:\Theta_z.
\label{thetaz}
\end{equation}
Note that if $\Delta_{z\zbar}\neq 0$, then $\Theta_z$ is {\em not\/} a
derivative of any function.

The presence of the one-loop K\"ahler metric in (\ref{del}) is not
the only source of violation of the integrability condition
(\ref{int}). This is due to the existence of anomalous couplings,
generated by the loops of massless particles, which give rise to
additional non-local terms in the effective action \cite{grs,dfkz}.
Space-time supersymmetry ensures however the validity of the relation
(\ref{int})  between the field-dependent gauge and axionic couplings
\cite{dkl2}.
In the following, by analyzing the integrability condition (\ref{int})
of the axionic couplings computed in string theory, we isolate the
local contribution of the Green-Schwarz term from the non-local,
field-theoretical, contribution. In this way, we rederive the formula
(\ref{gzz}) for the  one-loop K\"ahler metric as the universal (gauge
group-independent) part of $\Delta_{z\zbar}$. This method provides
therefore a direct determination of the Green-Schwarz term, whose
existence has been postulated before in the context of modular anomaly
cancellation \cite{dfkz}.

The one-loop string computation of the amplitude (\ref{thetaz}) yields
\cite{agn}:
\begin{equation}
\Theta_z~=~\frac{i}{{(2\pi)}^5}\int_{\Gamma}\frac{d^2\tau}{\tau_2}
\int{d^2\z}
\,\bar{\eta}(\bar{\tau})^{-2} \langle (Q^2 -
\frac{k}{4\pi\tau_2}){\cal T}_F(0)\Psi_z (\z) \rangle_{\rm odd},
\label{theta1}
\end{equation}
where $Q$ is the gauge charge operator. Taking the derivative
$\partial_{\zbar}\Theta_z$ amounts to inserting $\int{d^2
\xi}\overline{\Phi}_{\zbar}(\xi )$ [see (\ref{vzb})] inside the vacuum
average
(\ref{theta1}). In order to evaluate the difference (\ref{int}) we use
the relation $\overline{\Phi}_{\zbar} = \frac{1}{2\pi i} \oint 2{\cal T}_F
\psibar_{\zbar}$ and deform the contour of integration. The boundary term
vanishes due to the periodicity of the supercurrent  ${\cal T}_F$, and
we are left with two contributions. The first arises when the contour
encircles $\Psi_z(\z)$ converting it to $\Phi_z (\z)$. This is the same as
$\partial_z\Theta_{\zbar}$ and cancels in the difference (\ref{int}).
The second contribution comes when the contour encircles  ${\cal T}_F
(0)$ yielding an insertion of the stress-energy tensor of the internal
$N=2$ superconformal field theory. This reduces to a total derivative
with respect to the Teichm\"uller parameter $\tau$. We obtain
\begin{eqnarray}
\Delta_{z\zbar} &=& -\frac{i}{2{(2\pi)}^4}\int_{\Gamma} d^2\tau
\int{d^2\z}\, \bar{\eta}^{-2}\left\{\partial_{\tau}
\langle (Q^2 - \frac{k}{4\pi\tau_2}) \Psi_z(\z)\psibar_{\zbar}(0)
\rangle_{\rm odd}\right.\nonumber\\
 & & \hspace{5mm}\left.+~\partial_{\tau}\langle
\frac{k}{4\pi\tau_2}  \Psi_z(\z)\psibar_{\zbar}(0) \rangle_{\rm odd}
- \frac{k}{4\pi\tau_2} \partial_{\tau}\langle
\Psi_z(\z)\psibar_{\zbar}(0) \rangle_{\rm odd}\right\} ,
\label{int1}
\end{eqnarray}
where the total derivative term proportional to $k$ has been added
and subtracted, so that the first integrand is modular invariant.
Since it is a total derivative in $\tau$, its contribution to the
integral comes only  from the boundary of the moduli space, namely the
degeneration limit $\tau_2 \rightarrow \infty$. In this limit only the
$Q^2$ term contributes due to the presence of massless particles,
while the term proportional to $k$ vanishes due to the extra $\tau_2$
suppression. The final result is:
\begin{eqnarray}
\Delta_{z\zbar} &=&
\frac{-i}{2{(2\pi)}^4}\int_{\Gamma}d^2\tau \partial_{\tau} \int{d^2\z}\,
\bar{\eta}^{-2} \langle Q^2 \Psi_z(\z)\psibar_{\zbar}(0)
\rangle_{\rm odd}\nonumber\\ & & \hspace{5mm}+
\frac{k}{8{(2\pi)}^5}\int_{\Gamma}\frac{d^2\tau}{{\tau_2}^2} \int d^2{\z}
\,\bar{\eta}^{-2}\, \langle\Psi_z(\z)\psibar_{\zbar}(0)
\rangle_{\rm odd}\, .
\label{delta}
\end{eqnarray}

The above derivation of (\ref{delta}) is formal because the
intermediate expressions contain short-distance singularities. In the
Appendix we rederive (\ref{delta}) by using the momentum
regularization which respects conformal invariance.

The expression (\ref{delta}) for the non-harmonicity of gauge
couplings contains two parts. The group-dependent part proportional
to $Q^2$ was analyzed in Ref.\cite{agn} and was shown to reproduce
the field theory computation of one-loop anomalous graphs
involving massless particles \cite{dfkz,anom}. The universal part
({\em i.e}.\ the term proportional to $k$) should be identified with
the Green-Schwarz term. A comparison with (\ref{gzz}) shows that this
term is given by  the one loop correction to the K\"ahler metric, in
agreement with the field-theoretical expression (\ref{del}).

We can further relate the one-loop K\"ahler potential
to the quantity ${\rm Tr}F(-1)^F$ of the underlying $N=2$
superconformal theory \cite{cv}. In Ref.\cite{agn}, it was shown
that   \begin{equation}
\Delta~=~\frac{-i}{32\pi^2}
\int_{\Gamma}\frac{d^2\tau}{\tau_2}\bar{\eta}^{-2} {\rm
Tr}_R F(-1)^F (Q^2 - \frac{k}{4\pi\tau_2})
q^{L_0-\frac{c}{24}}\bar{q}^{\bar{L}_0-\frac{\bar{c}}{24}},
\label{trf}
\end{equation}
where $q=e^{2\pi i \tau}$ and the trace is over the Ramond sector
of the internal $N=2$ superconformal theory with left and right
central charges $c$ and $\bar{c}$, respectively. Note that the
integral (\ref{trf}) is infrared divergent. The coefficient of the
logarithmic divergence $\frac{d\tau_2}{\tau_2}$ is:
\begin {equation}
\frac{1}{32\pi^2}(-3{\rm Tr}Q_V^2 + {\rm Tr}Q_M^2),
\label{betaf}
\end{equation}
where the two terms are the contribution of gauginos and matter
fermions with $U(1)$-charges $F=\pm 3/2$ and $F=\pm 1/2$, respectively. The
expression (\ref{betaf}) coincides with the field-theoretical
one-loop $\beta$-function of gauge couplings in $N=1$ supersymmetric
Yang-Mills theory. Comparing the universal term in (\ref{trf}),
(\ref{delta}) and (\ref{gzz}), we find that the one-loop K\"ahler
potential is given by:
\begin{equation}
G^{(1)}~=~\frac{i}{16{(2\pi)}^3}
\int_{\Gamma}\frac{d^2\tau}
{\tau_2^2}\bar{\eta}^{-2} {\rm Tr}_R F(-1)^F
q^{L_0-\frac{c}{24}}\bar{q}^{\bar{L}_0-\frac{\bar{c}}{24}}.
\label{gzzf}
\end{equation}
It is worth noting that the same quantity was studied in Ref.\cite{cv},
in the massive case, where $F = F_L - F_R$, as a new kind of ``index"
for $N=2$ theories. At the conformal point one can take $F = F_L$ or
$F = F_R$  since then both $F_L$ and $F_R$ are conserved: in this case
one gets an expression which reminds of a loop space generalization
of the Ray-Singer torsion for K\"ahler manifolds.
Here, we
see that in the context of string theory this quantity plays the role
of the one-loop contribution to the physical  K\"ahler metric in the
effective $N=1$ supergravity theory.
\setcounter{equation}{0}

\section{$\beta$-functions of Yukawa couplings and threshold
corrections}

In this section, we compute the infrared divergence in the
one-loop correction to the matter metric. The coefficients of these
divergencies will be identified as the one-loop anomalous dimensions.
These also provide the beta-funtions of the physical Yukawa
couplings, which are entirely given by the wave function
renormalization factors, as already explained in Section 2. We also
discuss the threshold corrections.

The infrared divergence in the K\"ahler metric comes from the
$\tau_2\rightarrow\infty$ integration limit in the expression
(\ref{gzz}). In this limit the one-loop two-point correlator
degenerates to a sum over four-point functions on the sphere, and the
coefficient of the logarithmic divergence
$\frac{d\tau_2}{\tau_2}$ is:
\begin{equation}
G^{(1){\rm div}}_{z\zbar}\! = \frac{1}{8{(2\pi)}^4}
\displaystyle\lim_{\tau_2\rightarrow \infty}\frac{1}{\tau_2}\int
d\tau_1{\bar{\eta}}^{-2}{\rm Tr}(-1)^F
q^{L_0-\frac{c}{24}}\bar{q}^{\bar{L}_0-\frac{\bar{c}}{24}}\!\!\int\! d^2 x
\, x^{-\frac{1}{2}}\langle \bar{v}^R (0)  \Psi_z(x) \psibar_{\zbar}(1)
v^R (\infty) \rangle,
\label{div}
\end{equation}
where the trace is over all states of the internal conformal theory
in the Ramond sector with vertices $v^R$ and $\bar{v}^R$ (conjugate
vertices) in the $(-\frac{1}{2})$ ghost picture. The integration
of $x$ is in an annulus within radii $|q|^{\frac{1}{2}}$ and
$|q|^{-\frac{1}{2}}$ and the factor $x^{-\frac{1}{2}}$ comes from
the transformation of the torus coordinates to the annulus
coordinates (recall that $\Psi_z$ has dimension
$(\frac{1}{2},1)$). Inspection of (\ref{div}) shows that only the
massless states which have renormalizable interactions, {\em i.e}.\
gauginos and matter fermions contribute in the limit \cite{agn}.

The contribution of gauginos can be explicitly calculated using the
fact that the internal part of their vertices is $e^{\pm i
\frac{\sqrt{3}}{2} H}\frac{\bar{J}}{\sqrt{k}}$, where $\bar{J}$ is the
Ka\v{c}-Moody current and $H$ bosonizes the $N=2$ $U(1)$ current. The
result is:
\begin{equation}
-\frac{2}{k} C_2(R) G_{z\zbar}^{(0)},
\label{divg}
\end{equation}
where $C_2(R)$ is the quadratic Casimir of the representation $R$ to
which $z$ belongs.

Let us now consider the contribution of matter fields. One way to
extract the $\ln |q| \sim\tau_2$ behaviour of the integral over $x$
appearing in (\ref{div}) is to introduce non-zero external momenta and
look for poles in $s$ and $u$ in the limit $t \rightarrow 0$. We can
then follow the approach of Ref.\cite{agn} and relate the four-point
function appearing in (\ref{div}) to a physical four-point amplitude
involving matter scalars:
\begin{eqnarray}
x^{-\frac{1}{2}}\langle  V_{\bar{w}}^R (p_1,0)
V_{z}^{(-1)}(p_2,x) V_{\zbar}^{(-1)}(p_3,1)
V_{w}^R (p_4,\infty)\rangle & &
\nonumber\\ & &\hspace{-5cm}
 =~ -\frac{1}{u}\langle
V_{\bar{w}}^{(-1)}(p_1,0)  V^{(0)}_{z}(p_2,x)
V^{(0)}_{\zbar}(p_3,1) V_{w}^{(-1)}(p_4,\infty) \rangle
\label{A3}
\end{eqnarray}
where $u=p_2\cdot p_4$, and $w$ ($\bar{w}$) denote chiral
(anti-chiral) matter fields. A similar relation can be obtained when
$w$ and $\bar{w}$ are interchanged, with $s=p_1\cdot p_2$ replacing
$-u$ (the minus sign is due to the exchange of space-time fermions).
In the trace of (\ref{div}), one should include both the above terms
with a relative minus sign to take $(-1)^F$ into account. Using the
expressions for physical four scalar amplitudes given in
Ref.\cite{dkl1}, and taking the trace over $w$'s, we obtain
\begin{equation}
-\frac{2}{k} C_2(R) G_{z\zbar}^{(0)} + \frac{1}{2}
e^{G^{(0)}}W_{z z_1 z_2}G^{(0)z_1 \zbar_1}G^{(0)z_2
\zbar_2} \overline{W}_{\zbar \zbar_1 \zbar_2},
\label{divm}
\end{equation}
where $W_{z z_1 z_2}=\partial_z\partial_{z_1}\partial_{z_2}W$.
Combining the two contributions (\ref{divg}) and (\ref{divm}), we find
the following expression for the wave function renormalization:
\begin{equation}
-2\gamma_z = \frac{1}{32\pi^2}\left\{ -\frac{4}{k} C_2(R) +
\frac{1}{2} e^{G^{(0)}}W_{z z_2 z_3}G^{(0)z_2 \zbar_2}G^{(0)z_3
\zbar_3} \overline{W}_{\zbar_1 \zbar_2 \zbar_3}
G^{(0)z \zbar_1}\right\},
\label{divt}
\end{equation}
with no summation over $z$. In the expression (\ref{divt}), the
tree-level metric has been factorized out and for notational
simplicity we write diagonal elements only.
The above result reproduces directly the field-theoretical
one-loop beta-function of Yukawa couplings \cite{yuka}. In the
comparison one should take into account the normalization factor
appearing in the Yukawa interactions (\ref{ly}). Moreover,
the comparison with the field-theoretical anomalous dimensions
shows that the string computation implicitly uses a gauge in which the
superpotential remains unrenormalized.

When the logarithmic divergence in the $\tau_2$ integration is
regularized and compared to the field-theoretical $\overline{\rm DR}$
scheme, it is converted to $\ln\frac{M^2}{\mu^2}$, where $\mu$ is the
infrared cutoff corresponding to the low-energy scale and $M$ is the
string unification scale \cite{kap}. The remaining finite part gives
the string threshold corrections to wave function factors:
\begin{equation}
Y_z = \int_{\Gamma}\frac{d^2\tau}{\tau_2} \left\{
\frac{G^{(0)z \zbar_1}}{8{(2\pi)}^5\tau_2} \int
d^2{\z} \bar{\eta}^{-2}\, \langle\Psi_z(\z)\psibar_{\zbar_1}(0)
\rangle_{\rm odd}~+~2\gamma_z \right\}.
\label{thres}
\end{equation}
These corrections determine the boundary conditions of the physical
Yukawa couplings $\lambda_{ijk}$ at the unification scale:
\begin{equation}
\lambda_{ijk}(M) = \lambda_{ijk}^{\rm tree}\,
[1+g^2(Y_i + Y_j + Y_k)]^{-1/2},
\label{bcs}
\end{equation}
where $g=\sqrt{\langle l\,\rangle}$ is the string coupling constant.

We should point out that in contrast to the case of one-loop gauge
couplings, the infrared-divergent part of Yukawa couplings may be
in principle moduli-dependent, as seen from (\ref{divt}). In this case,
the moduli-dependence of threshold corrections cannot be
consistently defined.
\setcounter{equation}{0}

\section{Moduli metric in orbifold models}

In the last two sections we will consider some examples in orbifold
models. In particular we will compute the one-loop correction to
K\"ahler metric of untwisted moduli and matter. Here, we consider
the untwisted moduli, which we call generically $T$.\footnote{Our
arguments apply to both $T$-type and $U$-type moduli in the notation
of Ref.\cite{dkl2}.} The relevant  primary fields that appear in
(\ref{gzz}) are:
\begin{equation}
\Psi_T = \psi_L \dbar \xbar
\label{psi}
\end{equation}
where $X$ is a complex coordinate of an internal plane and $\psi_L$
is its left moving fermionic superpartner. $\psibar_{\tbar}$ is
the conjugate of $\Psi_T$.

In the orbifold models one has to sum over all sectors of boundary
conditions. The untwisted sector, which respects $N=4$ space-time
supersymmetry, gives vanishing contribution due to the internal
fermion zero-modes in the odd spin structure. In general there are
also sectors that preserve $N=2$ space-time supersymmetry, which appear
when one of the three internal planes is untwisted under the boundary
conditions. Such a sector could contribute if the modulus corresponds
to a deformation of the untwisted plane. In this case the moduli
vertices provide the two fermion zero-modes. Moreover $\xbar$'s in the
vertices are replaced by their classical solutions because the quantum
correlator being a total derivative does not contribute. After a
Poisson-resummation one can show that the result is:
\begin{equation}
G_{T\tbar}^{(1)} = \frac{1}{(T+\tbar)^2} {\cal I}~,~~~~~~~~~~~~
{\cal I} = \int_{\Gamma}\frac{d^2\tau}{{\tau_2}^2}
\partial_{\bar{\tau}}(\tau_2 Z) F(\bar{\tau}),
\label{n2}
\end{equation}
where $Z$ is the partition function:
\begin{equation}
Z = \sum_{p_L,p_R} q^{\frac{1}{2}p_L^2}
\bar{q}^{\frac{1}{2}p_R^2},
\label{z}
\end{equation}
with $p_L$ and $p_R$ the left and right momenta, respectively,
corresponding to the untwisted plane. Their dependence on $T$ and
$\tbar$ is well-known \cite{ver}. $F(\bar{\tau})$ is a meromorphic
modular form of weight $-2$  in $\tau$ and it does not depend on the
moduli $T$.

Using the arguments of Ref.\cite{ant}, we can derive a differential
equation for the quantity ${\cal I}$ of (\ref{n2}). The identity
\begin{equation}
\partial_T\partial_{\tbar} Z(\tau ,\bar{\tau})~=~
\frac{4\tau_2}{(T+\tbar )^2}\, \partial_{\bar{\tau}}
\partial_{\tau}(\tau_2 Z)
\label{lat}
\end{equation}
implies
\begin{equation}
\partial_T\partial_{\tbar} {\cal I}~=~\frac{4}{(T+\tbar )^2}
\int_{\Gamma}d^2\tau \{ \frac{i}{\tau_2}\partial_{\tau}
\partial_{\bar{\tau}}(\tau_2 Z) + \partial_{\tau}
\partial_{\bar{\tau}}^2(\tau_2 Z) \} F(\bar{\tau}).
\label{diff1}
\end{equation}
By doing a partial integration over $\tau$ and noting that the
surface terms vanish, we obtain the differential equation:
\begin{equation}
\partial_T\partial_{\tbar} {\cal I}~=~ \frac{2}{(T+\tbar )^2} {\cal I}.
\label{diff2}
\end{equation}
The general solution of the above equation is:
\begin{equation}
{\cal I} = [\frac{2}{(T+\tbar )} - \partial_T] f(T) +  c.c.\, ,
\label{diff3}
\end{equation}
where $f$ is a complex function of $T$.

{}From equation (\ref{n2}) it is clear that ${\cal I}$ is invariant under
the space-time duality transformations $T\rightarrow T+i$ and
$T\rightarrow 1/T$ \cite{sw}. This implies that $f$ must be a modular
function of weight 2 in $T$. Furthermore, by comparison with
(\ref{n2}) one can see that $f$ has no singularity inside the
fundamental domain, while at infinity it can at worst have a
powerlike singularity in $T$. This behaviour is inconsistent with
the invariance under $T \rightarrow T+i$ and analyticity.
Therefore $f$ must be holomorphic everywhere. Since a modular
function of weight 2 which is holomorphic everywhere does not
exist, $f$ must be zero, which in turn implies that ${\cal I}$ must
vanish. Hence, there are no corrections to the one-loop moduli metric
from $N=2$ sectors.

The remaining sectors preserve $N=1$ space-time supersymmetry. In this
case, all the internal coordinates are twisted and the integrand in
(\ref{gzz}) is proportional to the product of quantum correlators:
\begin{equation}
G^{(1)}_{T\tbar} =\frac{1}{8{(2\pi)}^5}
\int_{\Gamma}\frac{d^2\tau}{{\tau_2}^2}
\int d^2{\z} \,\bar{\eta}(\bar{\tau})^{-2}\,
\langle \psi_L (\z ) \bar{\psi}_L (0)\rangle_{\rm odd}
\langle\dbar \xbar(\bar{\z})\dbar
X(0)\rangle .
\label{n1}
\end{equation}
The bosonic correlator in (\ref{n1}) is a total derivative with
respect to $\bar{\z}$. The ${\z}$ integration can be performed after
regularizing the expression by cutting a disk around the origin, and
the result, in the $(g,h)$-twisted sector is $-\pi\partial_{\bar{\z}}\ln
\bar{\theta}_{(g,h)}({\bar{\z}})|_{\bar{\z}=0}$ times the partition
function of the internal superconformal theory in that sector.
Here ${\theta}_{(g,h)}$ denotes the odd $\theta$-function shifted
by the corresponding twists. Since the left moving bosonic and fermionic
contributions to the partition function cancel in the odd spin structure,
the resulting expression depends on $\bar{\tau}$ only. Summing over all
the twisted sectors one obtains:
\begin{equation}
G^{(1)}_{T\tbar} =\frac{-G^{(0)}_{T\tbar}}{16(2\pi)^4}
\int_{\Gamma}\frac{d^2\tau}{{\tau_2}^2} E(\bar{\tau})~,
\label{n1f}
\end{equation}
where we have explicitly extracted the tree-level metric
$G^{(0)}_{T\tbar}$ coming from the contractions of $X$'s and
$\psi_L$'s. The anti-analytic function $E(\bar{\tau})$ does
not depend on the moduli $T$. It is modular invariant in
$\bar{\tau}$ and has at most a simple pole singularity at $\bar{q}=0$.
It is therefore a priori proportional to the $j$ invariant up to an
additive constant.

In order to compute the $\tau$-integral in (\ref{n1f})
we can start from the following identity:
\begin{equation}
\int_{\Gamma}\frac{d^2\tau}{{\tau_2}^2}
E(\bar{\tau}) = -4
\int_{\Gamma}{d^2\tau} E(\bar{\tau})\partial_{\tau}\partial_{\bar{\tau}}
\ln(\tau_2 \bar{\eta}^2) .
\label{Elim}
\end{equation}
The integrand on the r.h.s. is a total derivative in $\tau$ of
a $(0,1)$-form, therefore
the integral receives contributions only from the boundary
$\tau_2\rightarrow \infty$ of
$\Gamma$. The $\tau_1$ integration then picks the constant term
in the $\bar{q}$-expansion of
\begin{equation}
4i E(\bar{\tau}) \partial_{\bar{\tau}}\ln[\bar{\eta}(\bar{\tau})].
\label{qzero}
\end{equation}
Actually this argument applies even if $E(\bar{\tau})$
has higher order poles at $\bar{q}=0$. Hence, the one-loop moduli
metric (\ref{n1f}) is proportional to the tree level metric, with a
coefficient determined by (\ref{qzero}):
\begin{equation}
G^{(1)}_{T\tbar} =-\frac{G^{(0)}_{T\tbar}}{4(2\pi)^4}
\displaystyle\lim_{\tau_2\rightarrow\infty} \int d\tau_1
E(\bar{\tau}) \partial_{\bar{\tau}}\ln[\bar{\eta}(\bar{\tau})].
\label{n1ff}
\end{equation}

It is very convenient to relate equation (\ref{n1ff}) with the
threshold corrections
to the gravitational coupling $\Delta^{\rm grav}$, defined as the
coefficient of the Gauss-Bonnet term.
In Ref.\cite{agn} it was shown that:
\begin{equation}
\Delta^{\rm grav,FT}_{T\tbar}=
\frac{1}{2{(2\pi)}^5}
\displaystyle\lim_{\tau_2\rightarrow\infty} \int d\tau_1
\partial_{\bar{\tau}}
\ln(\bar{\eta})\int d^2{\z}{\bar{\eta}}^{-2}
\langle \Psi_T (\z ) \psibar_{\tbar}(0)\rangle,
\label{grav}
\end{equation}
where FT denotes the ``field-theoretical''
part of the threshold correction.
In $N=1$ sectors, the $\z$-integral in (\ref{grav})
is identical with the one in (\ref{n1}),
which is equal to $-\pi G^{(0)}_{T\tbar}E(\bar{\tau})$.
Comparing the resulting expression with (\ref{n1ff}),
we conclude that the Green-Schwarz term is equal
to the contribution of $N=1$ sectors to the field-theoretical part of the
gravitational threshold correction.

For the moduli which have no $N=2$ subsectors,
as in the case of all untwisted moduli of $Z_3$ orbifold,
the Green-Schwarz
term is equal to the full field-theoretical contribution to the
gravitational
threshold correction. The latter has been calculated in \cite{agn} and
shown to be equal to the  group-dependent part of the
threshold correction to the $E_8$ gauge coupling.
This result is not surprising:
the universal  threshold
correction should cancel in this case against the gauge group-dependent
contribution of $N=1$ sectors in (\ref{delta}), since $N=1$
sectors do not contribute to the corresponding axionic coupling
$\Theta_T$ \cite{ant}. We have
verified in several orbifold examples that for such moduli the function
$E$ in (\ref{qzero}) is $2\pi$ times the $j$-function, which yields
the coefficient of $G^{(0)}_{T\tbar}$ in (\ref{n1f})
equal $-30/{16\pi^2}$; this is in agreement with the previous arguments.
The numerical evaluation of the coefficient in
(\ref{n1f}) can also be done for moduli which have $N=2$ subsectors
in various orbifolds and compared with the
coefficient of the Green-Schwarz term in the moduli-dependent
threshold corrections to gauge couplings \cite{dfkz,anom}.
In the case of a modulus corresponding
to a $Z_2$-twisted plane, the quantum correlator appearing in
(\ref{n1f}) is proportional to the $\z$-derivative of an even
$\theta$-function evaluated at $\z =0$. This is zero by the parity
properties of $\theta$-functions, implying that there is no one-loop
correction to the moduli metric, in agreement with the vanishing of the
Green-Schwarz term in this case.

As a result, the K\"ahler metric of the untwisted moduli in symmetric
orbifolds is renormalized at the one-loop order by a finite
multiplicative and calculable constant. Symmetric orbifolds are
particular examples of $(2,2)$ compactifications, which possess
$N=2$ world-sheet supersymmetry in both left and right moving
sectors. In this case, the gauge group is $E_6 \times E_8$ and the
matter fields transform as $27$ or $\overline{27}$ under $E_6$ and
they are in one-to-one correspondence with the moduli : $27$'s are
related to $(1,1)$ moduli and $\overline{27}$'s to $(1,2)$
moduli. Furthermore, the moduli metric is block-diagonal with
respect to these two types of moduli. An interesting consequence of
the  right-moving $N=2$ tree-level Ward-identities is the so-called
special geometry, which relates the tree-level moduli metric to the
Yukawa couplings \cite{dkl1}:
\begin{equation}
R^{(0)}_{a\bar{c}b\bar{d}} = G^{(0)}_{a\bar{c}}G^{(0)}_{b\bar{d}}
+ G^{(0)}_{a\bar{d}}G^{(0)}_{b\bar{c}} - e^{2G^{(0)}}
W_{abe}\overline{W}_{\bar{c}\bar{d}\bar{f}}G^{(0)e\bar{f}},
\label{sg}
\end{equation}
where $R^{(0)}_{a\bar{c}b\bar{d}}$ is the Riemann tensor of the
moduli metric $G^{(0)}$, and the above equation holds separately for
$(1,1)$ and $(1,2)$ moduli. Every term in (\ref{sg}) behaves
differently under global rescalings of the metric. Consequently,
our results imply that special geometry is in general violated
beyond the tree approximation. Note that from field-theoretical
point of view, special geometry is a consequence of $N=2$
space-time supersymmetry \cite{sg}. This is consistent with the fact
that orbifold sectors which preserve $N=2$ space-time supersymmetry
do not give rise to one-loop corrections to the moduli metric. As
a corollary, special geometry is preserved for moduli of planes
which are only $Z_2$-twisted.
\setcounter{equation}{0}

\section{\hspace*{-3mm}
Threshold corrections to Yukawa couplings in orbifold models}

In this section we consider the radiative corrections to the metric of
untwisted matter fields $z$. The relevant primary fields that appear in
(\ref{gzz}) have a form similar to (\ref{psi}) with $\dbar \xbar$
replaced by a right-moving fermion bilinear:
\begin{equation}
\Psi_z = \psi_L \psi_R \lambda ,
\label{psiz}
\end{equation}
where $\psi_{L,R}$ are left and right moving internal fermions, and
$\lambda$ is a right moving fermion generating the $SO(10)$ part
of $E_6$. As in the case of moduli, $N=4$ sectors give no
contribution, while $N=1$ sectors give contribution proportional
to the tree-level matter metric with a constant coefficient. Hence,
$N=1$ sectors yield moduli-independent corrections to the
physical Yukawa couplings (\ref{thres}). The moduli-dependent threshold
corrections arise therefore from $N=2$ sectors only, similarly to the
case of gauge couplings. Moreover, these sectors must leave invariant
the plane associated with the matter field $z$. The reason is that
the $z$ and $\zbar$ vertices should provide the two fermion zero
modes of the untwisted plane, otherwise the correlator vanishes.
As a result, the threshold correction $Y_z$ (\ref{thres}) may
depend only on the moduli of the plane associated with $z$. The
integral (\ref{gzz}) then becomes:
\begin{equation}
G^{(1)}_{z\zbar} =\frac{G^{(0)}_{z\zbar}}{8{(2\pi)}^5}
\int_{\Gamma}\frac{d^2\tau}{{\tau_2}^2} Z
\bar{\eta}(\bar{\tau})^{-2}\, \int d^2{\z}\, \langle
\psi_R (\bar{\z})\bar{\psi}_R(0) \lambda (\bar{\z})\lambda (0)
\rangle ,
\label{mat}
\end{equation}
where we extracted the tree-level matter metric from the various
contractions. $Z$ is the partition function (\ref{z}) of the untwisted
plane.

After summing over even spin structures of the right-moving
fermions, the quantum correlator in (\ref{mat}) becomes a
one-form in $\bar{\z}$ with a double pole at $\bar{\z}=0$.
Therefore, the normalized correlator can be written as
$-\partial_{\bar{\z}}^2 \ln\bar{\theta}_1$ up to an additive
constant. The $\z$-integration can then be performed with the
result:
\begin{equation}
G^{(1)}_{z\zbar} = -\frac{G^{(0)}_{z\zbar}}{8{(2\pi)}^3}
\int_{\Gamma}\frac{d^2\tau}{\tau_2} Z
\bar{\eta}(\bar{\tau})^{-2}
{\rm Tr} (\Gamma_z - \frac{1}{4\pi\tau_2})
\bar{q}^{\bar{L}_0-\frac{\bar{c}}{24}},
\label{mat1}
\end{equation}
where the operator $\Gamma_z$ present in the trace is the analog
of the $Q^2$ operator in the corresponding formula for gauge
couplings (\ref{trf}). Its eigenvalues on massless states contribute
to the coefficient of the infrared divergence which, in section 4, was
identified with the one loop anomalous dimension $\gamma_z$. Since
(\ref{mat1}) represents the contribution of the $N=2$ sectors only,
it is actually equal to $\frac{\hat{\gamma}_z}{ind}$. Here,
$\hat{\gamma}_z$ is the one-loop anomalous dimension coefficient of
the $z$-field in the corresponding $N=2$ space-time supersymmetric
theory with the orbifold defined by the little group of the unrotated
plane, and $ind$ is the index of this little group in the full
orbifold group. After subtracting this infrared divergence, we obtain
the threshold correction $Y_z$ (\ref{thres}) in a way explained in
Section 4.

In order to determine the moduli-dependence of $Y_z$ we follow the
method used in the case of gauge couplings \cite{ant}. We first
differentiate the integral in (\ref{mat1}) with respect to $T$ and
$\tbar$, and using the identity (\ref{lat}) we obtain:
\begin{equation}
\partial_T\partial_{\tbar}Y_z
{}~=~-\frac{1}{2{(2\pi)}^3}\frac{1}{(T+\tbar )^2}
\int_{\Gamma}d^2\tau \partial_{\bar{\tau}}
\partial_{\tau}(\tau_2 Z) \bar{\eta}(\bar{\tau})^{-2}
{\rm Tr}(\Gamma_z - \frac{1}{4\pi\tau_2})
\bar{q}^{\bar{L}_0-\frac{\bar{c}}{24}}.
\label{mat2}
\end{equation}
The second term in the r.h.s.\ of (\ref{mat2}), after partial
integration with respect to $\tau$, acquires the form of the
integral ${\cal I}$ appearing in (\ref{n2}) and vanishes by the same
reasoning. The first term in the r.h.s. of (\ref{mat2}) is a
total derivative with respect to $\tau$ and therefore receives
contributions from massless states only. The situation becomes
similar to the case of gauge couplings when one considers the
violation of the integrability condition for $\Theta$-angles. Hence,
one can replace $\bar{\eta}^{-2}{\rm Tr}\Gamma_z
\bar{q}^{\bar{L}_0-\frac{\bar{c}}{24}}$ by
$\frac{\hat{\gamma}_z}{ind}$. The boundary integration in (\ref{mat2})
then gives:
\begin{equation}
\partial_T\partial_{\tbar}Y_z ~=~-\frac{2\hat{\gamma}_z}{ind}
\frac{1}{(T+\tbar )^2}. \label{mat3}
\end{equation}
By integrating the differential equation (\ref{mat3}) and using duality
invariance, one obtains the following formula for the threshold
corrections to the wave function factors $Y_z$ of untwisted matter
fields:
\begin{equation}
Y_z~=~ \frac{2\hat{\gamma}_z}{ind}
\ln [|\eta(iT)|^4(T+\tbar)] + y_z,
\label{mat4}
\end{equation}
where $y_z$ is a moduli-independent constant.

The coefficients $\hat{\gamma}_z$ are anomalous dimensions of
untwisted scalar fields $z$ in an $N=2$ supersymmetric orbifold. These
fields belong to $N=2$ vector supermultiplets. For instance if $z$
is in the 27 representation of $E_6$, in the corresponding $N=2$
theory it belongs to a gauge vector multiplet of $E_7$. Consequently,
its anomalous dimension $\hat{\gamma}_z =-\hat{b}_z^i/2$, where
$\hat{b}_z^i$ is the corresponding beta function coefficient
\cite{dkl2} of any subgroup that transforms $z$ non-trivially; $i$
denotes the plane associated with $z$.

As an example consider the Yukawa coupling between three untwisted
27's of $E_6$. At the tree-level this coupling is:
\begin{equation}
\lambda_{ijk}^{\rm tree} = \frac{g}{\sqrt{2}}W_{ijk},
\label{yukun}
\end{equation}
where $W_{ijk}$ are constants which are non zero only if the three
27's are associated with three different planes. In this case
(\ref{mat4}) combined with (\ref{bcs}) gives the boundary condition:
\begin{equation}
\lambda_{ijk}(M) = \frac{g_{E_6}(M)}{\sqrt{2}}W_{ijk}\,
[1+g_{E_6}^2(M)y_{ijk}]^{-1/2},
\label{yukbc}
\end{equation}
where $y_{ijk}$ are moduli-independent constants, and $g_{E_6}(M)$ is
the one-loop $E_6$ gauge coupling \cite{dkl2,ant}:
\begin{equation}
\frac{1}{g_{E_6}^2(M)} = \frac{1}{g^2} -
\sum_{i} \frac{\hat{b}_{E_6}^i}{(ind)^i}\ln
[(|\eta(iT^i)|^4(T^i+\tbar^i)] ~+~ c_{E_6},
\label{e6}
\end{equation}
with $c_{E_6}$ being another moduli-independent constant.
As a result, the boundary relation between the untwisted Yukawa
couplings and the $E_6$ gauge coupling at the unification scale does
not receive any moduli-dependent corrections at the one loop level.\\[1cm]
{\bf Acknowledgements}

The authors acknowledge the hospitality of the Department of Physics at
Northeastern University, ICTP, and of the Centre de Physique
Th\'eorique at Ecole Polytechnique. T.R.T.\ thanks CERN Theory
Division for its hospitality during completion of this work and
acknowledges useful discussions with S. Ferrara and J. Louis.

\begin{flushleft}
{\large\bf Appendix}\end{flushleft}
\renewcommand{\theequation}{A.\arabic{equation}}
\renewcommand{\thesection}{A.}
\setcounter{equation}{0}

Here, we rederive (\ref{delta}) by using the momentum regularization
of short-distance singularities. The derivative $\partial_{\zbar}
\Theta_z$ is given by:
\begin{eqnarray}\hspace{-3mm}
i\epsilon^{\mu\nu\lambda\rho}p_{1\lambda}p_{2\rho}\:
\partial_{\zbar}\Theta_z  &=&
\nonumber\\ & &\hspace*{-3cm}
\int_{\Gamma}\frac{d^2\tau}{\tau_2}\lim_{p_4\rightarrow 0}
\int\prod_{i=1}^{4} d^2\z_i
\left\langle V_A^{\mu}(p_1,\z_1)\, V_A^{\nu}(p_2,\z_2)\,
V_z^{(-1)}(p_3,\z_3)\, V_{\zbar} (p_4,\z_4)\,
{\cal T}_F(0)\right\rangle_{\rm odd}^{\rm 1PI},
\label{reg}
\end{eqnarray}
where the vertices are given in (\ref{vz}) and by:
\begin{equation}
V_{\zbar}^{(-1)}(p,\z) ~=~ :\!\psibar_{\zbar} e^{ip\cdot
\!X}\!:~,\label{vzb1}\vspace{-3mm}
\end{equation}
\begin{equation}
V_A^{\mu}(p,\z) ~=~ :\!\bar{J}(\bar{\z})\, (\partial X^{\mu}+
ip\!\cdot\!\psi\,\psi^{\mu}) e^{ip\cdot \!X}\!:\vspace{1mm}
\label{va}
\end{equation}
with $\bar{J}$ being the Ka\v{c}-Moody current.
The one-particle irreducible (1PI) amplitude
(\ref{reg}) is obtained from the full amplitude by subtracting the
reducible diagram involving the exchange of the
antisymmetric tensor between the tree-level $b$-$A$-$A$ and the
one-loop $b$-$z$-$\zbar$ vertices. In the $p_4\rightarrow 0$
limit, this subtraction procedure is gauge invariant since the
intermediate antisymmetric tensor enters on-shell. More
explicitly, in the Feynman gauge for the antisymmetric tensor this
amounts to subtracting from the full amplitude the following
expression:
\begin{equation}
\langle A^{\mu}(p_1)A^{\nu}(p_2)z(p_3)\zbar(p_4)\rangle_{\rm
odd}^{\rm 1PR} =
\frac{k}{2}\epsilon^{\mu\nu\lambda\rho}(p_1 -
p_2)_{\lambda} (p_3 - p_4)_{\rho}
G^{(1)}_{z\zbar}\, ,
\label{red}
\end{equation}
which is manifestly gauge invariant in the limit of zero $p_3$ or
$p_4$. In the string computation of the four-point
function the above expression appears through so-called contact terms.
For example one can consider the term in the four point string
amplitude involving the space time part of ${\cal T}_F$.
Contracting ${\cal T}_F$ with $\psi^{\mu}$ from $V_A^{\mu}$ and
$e^{ip_2\cdot \!X}$ in $V_A^{\nu}$, we find a singularity of the
form $1/(\z -\z_1)(\z -\z_2)$. Furthermore the contraction of the
Ka\v{c}-Moody currents yields a singularity of the form
$k/(\bar{\z}_1-\bar{\z}_2)^2$. To start with, this term is quartic
in momenta but the integration of $\z_1$ and $\z_2$ gives a pole in
$p_1\cdot p_2$ giving rise again to a quadratic term in momenta.
Combining the term coming from interchanging the two gauge fields,
a straightforward calculation yields the r.h.s.\ of (\ref{red}) with
$G^{(1)}_{z\zbar}$ given by (\ref{gzz}). There are also other contact
terms in the string amplitude. By contracting ${\cal T}_F$ with
$p_1\cdot \psi$ instead of $\psi^{\mu}$ in the above, one gets
additional terms that can be interpreted as the ones coming from the
second term in (\ref{ff}) involving the gauge topological current
$\omega^{\mu}$. As expected, this term is also not separately gauge
invariant. However, as one can check explicitly, after combining all
the contact terms, only the kinematic form
$\epsilon^{\mu\nu\lambda\rho} p_{1\lambda}p_{2\rho}$ survives
which is gauge invariant. To proceed further,
$\partial_z\Theta_{\zbar}$ is defined in a similar way with the
vertex of $\zbar$ appearing in the $(-1)$-ghost picture and the
limit $p_3\rightarrow 0$.

In order to evaluate the difference (\ref{int}) we can proceed as in
\cite{agn}. Expressing $V_{\zbar} =\frac{1}{2\pi i}\oint 2{\cal T}_F
V_{\zbar}^{(-1)}$  in (\ref{reg}) and deforming the contour
integration one finds two contributions. The first arises when the
contour encircles  ${\cal T}_F$ in (\ref{reg}) yielding an
insertion of the energy-momentum tensor and reduces to a total
derivative with  respect to the Teichm\"uller parameter $\tau$:
\begin{equation}
\frac{-i}{2{(2\pi)}^4}\int_{\Gamma}d^2\tau \partial_{\tau} \int{d^2\z}\,
\bar{\eta}(\bar{\tau})^{-2} \langle (Q^2 -
\frac{k}{4\pi\tau_2})\Psi_z (\z)\psibar_{\zbar}(0) \rangle_{\rm odd}.
\label{tb}
\end{equation}
The second contribution comes when the contour encircles
$V_z^{(-1)}$ in (\ref{reg}) converting it to $V_z$. This is the
same as the expression for $\partial_z\Theta_{\zbar}$ except for
the different momentum limit, namely the limit $p_4\rightarrow
0$. Therefore the computation of $\Delta_{z\zbar}$ reduces to
evaluating the difference between the two limits: $p_4\rightarrow 0$
and $p_3\rightarrow 0$ of the 1PI part of the four point function
$\left\langle V_A^{\mu}(p_1) V_A^{\nu}(p_2) V_z(p_3)
V_{\zbar}^{(-1)}(p_4){\cal T}_F\right\rangle_{\rm odd}$. Now the
one-particle irreducible part of the amplitude is given by the
difference of the full corresponding four-point function and the
reducible part (\ref{red}). As stated earlier the full four-point
function is gauge invariant and comes with the kinematic
structure  $\epsilon^{\mu\nu\lambda\rho}p_{1\lambda}p_{2\rho}$ which
is independent of the two momentum limits and therefore vanishes in
the difference. On the other hand the reducible part gives
equal contribution with opposite signs in these two limits as can be
easily seen from (\ref{red}). Thus in the difference they add up.
Combining both these contributions, we reproduce equation
(\ref{delta}). This derivation makes it clear that the
group-dependent contribution to $\Delta_{z\zbar}$, which corresponds to
the first term of (\ref{delta}), comes from the irreducible part of
the amplitude (\ref{reg}), whereas the universal contribution
proportional to the Ka\v{c}-Moody level arises from the reducible
diagrams.

\end{document}